\documentclass[aps,pra,twocolumn,reprint]{revtex4}
\usepackage{amsfonts, amssymb, amsmath,graphicx}
\usepackage{subfigure}
\usepackage{multirow}
\usepackage{dcolumn}
\usepackage{bm}
\usepackage{color}

\begin{document}

\title{Quantum Hall effect in momentum space}

\author{Tomoki Ozawa}
\author{Hannah M. Price}
\author{Iacopo Carusotto}
\affiliation{INO-CNR BEC Center and Dipartimento di Fisica, Universit\`a di Trento, I-38123 Povo, Italy}%

\date{\today}

\def\simge{\mathrel{%
         \rlap{\raise 0.511ex \hbox{$>$}}{\lower 0.511ex \hbox{$\sim$}}}}
\def\simle{\mathrel{
         \rlap{\raise 0.511ex \hbox{$<$}}{\lower 0.511ex \hbox{$\sim$}}}}
\newcommand{\feynslash}[1]{{#1\kern-.5em /}}
\newcommand{\tom}[1]{{\color{red} #1}}
\newcommand{\iac}[1]{{\color{blue} #1}}

\begin{abstract}
We theoretically discuss a momentum-space analog of the quantum Hall effect, which could be observed in topologically nontrivial lattice models subject to an external harmonic trapping potential. In our proposal, the Niu-Thouless-Wu formulation of the quantum Hall effect on a torus is realized in the toroidally shaped Brillouin zone. In this analogy, the position of the trap center in real space controls the magnetic fluxes that are inserted through the holes of the torus in momentum space. We illustrate the momentum-space quantum Hall effect with the noninteracting trapped Harper-Hofstadter model, for which we numerically demonstrate how this effect manifests itself in experimental observables. Extension to the interacting trapped Harper-Hofstadter model is also briefly considered. We finally discuss possible experimental platforms where our proposal for the momentum-space quantum Hall effect could be realized.
\end{abstract}

\maketitle

\section{Introduction}

The discovery of the quantum Hall effect revealed the dramatic role that topological properties of quantum states can have on observable quantities. In the quantum Hall effect, the transverse Hall conductivity of a two-dimensional (2D) electron gas is precisely quantized in the presence of a strong, static, and perpendicular magnetic field~\cite{Klitzing:1980,Tsui:1982}. The first step towards a theoretical understanding of this effect was put forward by Laughlin~\cite{Laughlin:1981}. By considering a 2D electron gas confined to the curved surface of a cylinder, Laughlin found that the quantized Hall conductivity was related to the response of the system to an insertion of a magnetic flux through the ends of the cylinder. Soon afterwards, Thouless, Kohmoto, Nightingale, and den Nijs (TKNN)~\cite{Thouless:1982} discovered that the quantum Hall response could be expressed in terms of a topological invariant of the energy band, the so-called Chern number, opening up the new field of topological insulators and superconductors in solid-state physics~\cite{Hasan:2010, Qi:2011}.

A few years after the TKNN paper, an elegant formulation of the quantum Hall effect was proposed by Niu, Thouless, and Wu~\cite{Niu:1985, Niu:1987}. By imagining the 2D electron gas confined to the surface of a torus, they found that the Hall conductance could be written in terms of the response of the system to twists of the boundary condition around the torus, or, equivalently, to the insertion of magnetic fluxes through the holes of the torus. This formalism has the advantage over other approaches that it is not restricted to non-interacting, clean systems, but can also describe the many-body quantum Hall effect in the presence of interactions and disorder. Thanks to its wide applicability, the Niu-Thouless-Wu formulation of the quantum Hall effect has therefore been used extensively as a powerful tool to theoretically understand both the integer and fractional quantum Hall effects~\cite{Fradkin:Book}. The quantum Hall effect on a torus is also itself an important subject of study as a system in a fractional quantum Hall state is generally characterized by a topologically-protected ground-state degeneracy when the underlying geometry is multiply connected~\cite{Oshikawa:2006}. Unfortunately, the experimental difficulty of preparing a quantum Hall sample in the toroidal geometry, which is essential for the Niu-Thouless-Wu formulation, has so far hindered a direct experimental realization of this {\it gedankenexperiment}.

In all of the works quoted above, as well as in most of the literature, the quantum Hall effect was studied in the usual {\em real-space} case, i.e., where the Hall current is an electric current that flows in real space along the physical edges of the sample and where the conductance in response to a real-space electric field is quantized to an integer value.

\begin{figure}[htbp]
\begin{center}
\includegraphics[width=8.0cm]{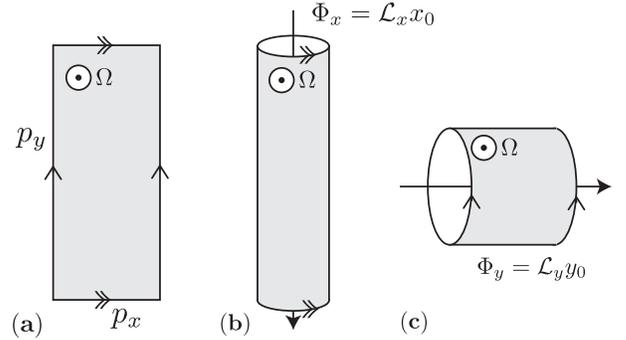}
\caption{The analog magnetic fluxes that are inserted through the holes of the toroidally-shaped Brillouin zone, whose length in the $p_x$ direction is $\mathcal{L}_x$ and in the $p_y$ direction is $\mathcal{L}_y$. In this analogy, the Berry curvature $\Omega (\mathbf{p})$ acts as an effective magnetic field in momentum space. The trap center is located at $(x_0, y_0)$. (a) The toroidal structure of the Brillouin zone. (b) The flux corresponding to $x_0 \neq 0$. (c) The flux corresponding to $y_0 \neq 0$.}
\label{insertingflux}
\end{center}
\end{figure}

In this paper, we show that a generalized quantum Hall effect can also occur in momentum space. A 2D topologically-nontrivial model in momentum space is realized by applying a harmonic confining potential to an underlying 2D topologically-nontrivial real-space model. The Berry curvature of the underlying energy bands acts as an effective magnetic field in momentum space and the underlying band dispersion acts as a periodic potential in momentum space, while the confining potential provides the momentum-space kinetic energy~\cite{Price:2014,Ozawa:2015}. The global connectivity of momentum space is also set by the Brillouin zone, which naturally maps onto a torus as the quasi-momentum is periodic with respect to the reciprocal lattice vectors. Importantly, it turns out that moving the center of the harmonic trap through the two-dimensional real space corresponds to the insertion of magnetic fluxes through the two holes of the torus in momentum space, as depicted in Fig.~\ref{insertingflux} and further described in the body of the paper. According to the Niu-Thouless-Wu formalism, the momentum-space Hall conductivity is then related to the real-space displacement of the wave function from the trap center when the latter is moved in real space so as to vary the momentum-space magnetic flux. As a result, this proposal of the {\em quantum Hall effect in momentum space} provides the first realistic system for which the quantum Hall effect on a torus can be observed in experiments. We note that in terms of the formalism of the dynamical quantum Hall effect developed in~\cite{Gritsev:2012}, our momentum-space quantum Hall effect can be interpreted as a dynamical quantum Hall effect where the dynamical parameter is the position of the trap center.

In Sec.~\ref{reviewniu}, we briefly review the Niu-Thouless-Wu formalism of the quantum Hall effect, before reviewing the effective momentum-space theory of a particle in topologically nontrivial models in the presence of an external harmonic trap in Sec.~\ref{msf}. Based on these concepts, we then proceed in Sec.~\ref{msqhe} to present the momentum-space quantum Hall effect for general momentum-space topological models. In Sec.~\ref{thhm}, the general idea is applied to the specific case of a trapped non-interacting Harper-Hofstadter model, for which we numerically confirm the expected features of the quantum Hall effect in momentum space. We extend our discussion to include the effect of weak interactions in this model in Sec.~\ref{sec:interaction}. Finally, in Sec.~\ref{sec:discussion} we discuss possible experimental setups based on ultracold gases and photonics where the momentum-space quantum Hall effect could be observed in the near future, before drawing conclusions and presenting future perspectives in Sec.~\ref{sec:conclusion}.

\section{The Niu-Thouless-Wu formalism of the real-space quantum Hall effect}
\label{reviewniu}

We first review the Niu-Thouless-Wu formalism of the usual quantum Hall effect in real sapce~\cite{Niu:1985, Niu:1987}. We consider a two-dimensional system of $N$ particles whose Hamiltonian is
\begin{align}
	\mathcal{H}
	=
	\sum_{a=1}^N
	\left(
	\frac{[\mathbf{p}_a - \mathbf{A}(\mathbf{r}_a)]^2}{2m} + V(\mathbf{r}_a)
	\right)
	+
	\mathcal{H}_\mathrm{int}, \label{firsth}
\end{align}
where $m$ is the mass of a particle, $\mathbf{p}_a$ is the momentum of a particle labeled $a$, and $\mathbf{A}(\mathbf{r}_a)$ and $V(\mathbf{r}_a)$ are the vector potential and the scalar potential of the particle at position $\mathbf{r}_a$. We have set the charge of a particle to be unity. The term $\mathcal{H}_\mathrm{int}$ takes into account inter-particle interactions.

When the system is in a many-body eigenstate $|\alpha \rangle$ of the Hamiltonian (\ref{firsth}), the Hall conductivity in the presence of an additional external electric field is given by the Kubo formula:
\begin{align}
	\sigma_{xy}
	=
	\frac{-i}{L_x L_y}\sum_{\beta \neq \alpha}
	\frac{
	\langle \alpha | J_x | \beta \rangle \langle \beta | J_y | \alpha \rangle
	-
	\langle \alpha | J_y | \beta \rangle \langle \beta | J_x | \alpha \rangle
	}{(E_\alpha - E_\beta)^2}
\end{align}
where $L_x$ and $L_y$ are the lengths of the system in the $x$ and $y$ directions, respectively, and we have set $\hbar = 1$. The sum is over all many-body eigenstates $|\beta\rangle$ different from $|\alpha\rangle$, where $E_\alpha$ and $E_\beta$ are the energies of the states $|\alpha\rangle$ and $|\beta\rangle$, respectively. The $\mu$th component of the total current operator $\mathbf{J}$ is
\begin{align}
	J_\mu
	=
	\sum_{a=1}^N \frac{p_{a,\mu} - A_\mu (\mathbf{r}_a)}{m}. \label{totalcurrent}
\end{align}

Now, we assume that the particles are confined to move on the surface of a torus.
In the presence of a non-zero magnetic field, the vector potential $\mathbf{A}(\mathbf{r})$ cannot be made periodic over the torus, and instead satisfies the following boundary conditions~\cite{Fradkin:Book}:
\begin{align}
	A_x (x, y + L_y) = A_x (x,y) + \partial_x \phi_y (x,y), \notag \\
	A_y (x + L_x, y) = A_y (x,y) + \partial_y \phi_x (x,y),
\end{align}
with transition functions $\phi_x$ and $\phi_y$ which are uniquely defined up to a constant additive factor.

Correspondingly, also the wave function $\Psi (\{ \mathbf{r}_a\}) = \langle \{ \mathbf{r}_a \} | \alpha \rangle$ is not periodic but satisfies
\begin{align}
	\Psi(\{x_a + L_x, y_a\}) &= e^{i\sum_a \left[ \phi_x (\{x_a, y_a\}) + \theta_x\right]} \Psi(\{x_a, y_a\}), \notag \\
	\Psi(\{x_a, y_a + L_y\}) &= e^{i\sum_a \left[ \phi_y (\{x_a, y_a\}) + \theta_y\right]} \Psi(\{x_a, y_a\}),
	\label{twistbc}
\end{align}
where the additional constants $\theta_x$ and $\theta_y$ can be arbitrarily chosen so as to provide the desired ``twist angles" in the boundary conditions.

The physical meaning of these twist angles can be better understood by performing the gauge transformation,
\begin{align}
	\tilde{\Psi}(\{x_a, y_a\})
	&\equiv
	e^{-i\theta_x \sum_a x_a/L_x -i\theta_y \sum_a y_a /L_y} \Psi(\{x_a, y_a\}), \notag \\
	\tilde{\mathbf{A}}(\mathbf{r}_a)
	&\equiv
	\mathbf{A}(\mathbf{r}_a) - \boldsymbol{\theta_L},
\end{align}
where $\boldsymbol{\theta_L} \equiv (\theta_x/L_x, \theta_y/L_y)$.
With respect to this transformed wave function $\tilde{\Psi}$, the boundary conditions no longer involve the twist $\boldsymbol{\theta_L}$, which now appears as a constant term in the magnetic vector potential. As in Laughlin's \textit{gedankenexperiment}~\cite{Laughlin:1981,Yoshioka:Book}, when the underlying manifold is a torus, a constant shift in vector potential corresponds to inserting a solenoid flux through the two holes of the torus with a magnetic flux $\theta_x$ and $\theta_y$, respectively.
Thus, a system with a twisted boundary condition~(\ref{twistbc}) without a flux through the torus is related via a gauge transformation to a system without a twist in the boundary condition but with a flux through the torus; the two pictures are fully equivalent.

Let us denote by $|\tilde{\alpha}\rangle$ and $|\tilde{\beta}\rangle$ the many-body eigenstates corresponding in the transformed gauge to the states $|\alpha\rangle$ and $|\beta\rangle$ in the original gauge. Then, in the transformed gauge the Hall conductivity is
\begin{align}
	\sigma_{x y}
	=
	\frac{-i}{L_x L_y}\sum_{\tilde{\beta} \neq \tilde{\alpha}}
	\frac{
	\langle \tilde{\alpha} | \tilde{J}_x | \tilde{\beta} \rangle \langle \tilde{\beta} | \tilde{J}_y | \tilde{\alpha} \rangle
	-
	\langle \tilde{\alpha} | \tilde{J}_y | \tilde{\beta} \rangle \langle \tilde{\beta} | \tilde{J}_x | \tilde{\alpha} \rangle
	}{(E_\alpha - E_\beta)^2}, \label{notyet}
\end{align}
where
\begin{align}
	\tilde{J}_\mu
	=
	\sum_{a=1}^N \frac{p_{a,\mu} - \tilde{A}_\mu (\mathbf{r}_a)}{m}
	=
	L_\mu
	\frac{\partial \tilde{\mathcal{H}}}{\partial \theta_\mu}, \label{tildejmu}
\end{align}
and the Hamiltonian reads as
\begin{align}
	\tilde{\mathcal{H}}
	=
	\sum_{a=1}^N
	\left(
	\frac{[\mathbf{p}_a - \tilde{\mathbf{A}}(\mathbf{r}_a)]^2}{2m} + V(\mathbf{r}_a)
	\right)
	+
	\mathcal{H}_\mathrm{int}. \label{hamshiftreal}
\end{align}
Using the relation 
\begin{align}
	\langle \tilde{\alpha} | \tilde{J}_\mu | \tilde{\beta} \rangle
	=
	L_\mu (E_\alpha - E_\beta)
	\left\langle \frac{\partial \tilde{\alpha}}{\partial \theta_\mu}\right|
	\left. \tilde{\beta} \right\rangle, \label{formula1}
\end{align}
whose proof is sketched in the footnote~\footnote{Taking the derivative of the both sides of $\tilde{\mathcal{H}} |\tilde{\beta}\rangle = E_\beta |\tilde{\beta}\rangle$, and applying $\langle \tilde{\alpha}|$ from the left, one obtains 
$\langle \tilde{\alpha} | \frac{\partial \mathcal{H}}{\partial \theta_\mu} | \tilde{\beta}\rangle = -(E_\alpha - E_\beta) \langle \tilde{\alpha}| \frac{\partial \tilde{\beta}}{\partial \theta_\mu} \rangle$.
Multiplying the both sides by $L_\mu$ and noticing that $\langle \tilde{\alpha}| \frac{\partial \tilde{\beta}}{\partial \theta_\mu} \rangle = -\langle \frac{\partial \tilde{\alpha}}{\partial \theta_\mu}| \tilde{\beta}\rangle$, one obtains~(\ref{formula1}).}, 
and the completeness relation that $\sum_{\tilde{\beta}} |\tilde{\beta}\rangle \langle \tilde{\beta}|$ is equal to the identity, the expression (\ref{notyet}) simplifies to
\begin{align}
	\sigma_{xy}
	&=
	-i
	\left(
	\left\langle \frac{\partial \tilde{\alpha}}{\partial \theta_x}\right|
	\left. \frac{\partial \tilde{\alpha}}{\partial \theta_y} \right\rangle
	-
	\left\langle \frac{\partial \tilde{\alpha}}{\partial \theta_y}\right|
	\left. \frac{\partial \tilde{\alpha}}{\partial \theta_x} \right\rangle
	\right)
	\notag \\
	&\equiv
	-\Omega (\theta_x, \theta_y). \label{rsxy}
\end{align}
This is the Berry curvature defined in the parameter space spanned by $(\theta_x, \theta_y)$. As shown in~\cite{Niu:1985, Niu:1987}, the conductance $\sigma_{xy}$ no longer depends on the boundary conditions $\theta_x$ and $\theta_y$ in the thermodynamic limit. Then we can take an average over $\theta_x$ and $\theta_y$, which results in
\begin{align}
	\sigma_{xy}
	=
	-\frac{1}{(2\pi)^2}\int_0^{2\pi}d\theta_x \int_0^{2\pi}d\theta_y\, \Omega (\theta_x, \theta_y)
	=
	-\frac{1}{2\pi}\mathcal{C},
\end{align}
where $\mathcal{C}$ is the Chern number in $(\theta_x, \theta_y)$ space: depending on the nature of the state of interest, this Chern number $\mathcal{C}$ can either be an integer when the state is non-degenerate and gapped from other states (the integer quantum Hall effect), or can take a fractional value when the state is degenerate and gapped (the fractional quantum Hall effect).

This argument for the quantization of the Hall conductance relies on the assumption that we have a toroidal geometry, and that the magnetic flux can be inserted through the holes of the torus or, equivalently, that the boundary condition of the system can be adjusted. The torus geometry and the flux insertion are difficult to experimentally realize in the ordinary setup of the quantum Hall effect in real space. Therefore, although the Hall conductance has been measured in a simpler cylindrical (Corbino disk) geometry~\cite{Dolgopolov:1992}, the measurement of the Hall conductance in the full toroidal geometry is lacking, and thus the Niu-Thouless-Wu formulation has so far only been used for theoretical purposes. 

In contrast, the main result of this paper is that both a toroidal geometry and a flux insertion mechanism can be realized in momentum space, resulting in the quantization of the analog of the Hall conductivity in momentum space. In the next section, we will follow~\cite{Price:2014,Ozawa:2015}  and review how topologically non-trivial momentum-space models can arise from harmonically trapped topological models in real space. In particular, we will show how a controllable magnetic flux can be inserted through the holes of the momentum-space torus by simply displacing the center of the harmonic trap in real space.

\section{Effective Momentum-space Hamiltonian}
\label{msf}

In this section, we review the momentum-space theory of a particle in a topologically nontrivial model with a harmonic confinement potential.
We consider a two-dimensional real-space lattice model with a harmonic confining potential, described by the following single-particle Hamiltonian:
\begin{align}
	\mathcal{H}
	=
	\mathcal{H}_0 + \frac{\kappa}{2} \left\{(x-x_0)^2 + (y-y_0)^2 \right\}, \label{hamoriginal}
\end{align}
where $\mathcal{H}_0$ is a periodic lattice Hamiltonian, and $\kappa$ is the strength of the harmonic confinement. The harmonic potential is centered at a position $\mathbf{r}_0 \equiv (x_0, y_0)$, which need not coincide with a lattice site.

We assume that $\mathcal{H}_0$ has $q$ bands, and that these have nontrivial geometry, namely, the Berry connection $\boldsymbol{\mathcal{A}}_{n,n^\prime} \equiv i\langle n,\mathbf{p}|\nabla_\mathbf{p}|n^\prime, \mathbf{p}\rangle$ is generally nonzero. Here, $|n,\mathbf{p}\rangle$ is the Bloch state of $\mathcal{H}_0$ labeled by the band index $n$ and the quasimomentum $\mathbf{p}$ and satisfying $\mathcal{H}_0 e^{i\mathbf{p}\cdot \mathbf{r}}|n,\mathbf{p}\rangle = E_n (\mathbf{p}) e^{i\mathbf{p}\cdot \mathbf{r}}|n,\mathbf{p}\rangle$, where $E_n (\mathbf{p})$ is the band dispersion of the $n$-th band. In order to derive an effective model in momentum space, we expand the wave function $|\Psi\rangle$ in the Bloch state basis as $|\Psi\rangle = \sum_{n,\mathbf{p}}\psi_{n,\mathbf{p}} e^{i\mathbf{p}\cdot \mathbf{r}}|n,\mathbf{p}\rangle$.

Assuming for the moment that the trap center is at the origin ($x_0 = y_0 = 0$), the time evolution of the momentum-space wave function $\psi_{n,\mathbf{p}}$ obeys~\cite{Price:2014}
\begin{align}
	i\frac{\partial}{\partial t}\hat{\psi}_\mathbf{p}(t)
	=
	\hat{E}(\mathbf{p}) \hat{\psi}_\mathbf{p} (t)
	+
	\frac{\kappa}{2}
	\left( I_{q} i\nabla_\mathbf{p} + \overleftrightarrow{\boldsymbol{\mathcal{A}}}(\mathbf{p}) \right)^2
	\hat{\psi}_\mathbf{p}(t),
\end{align}
where $\hat{\psi}_\mathbf{p}(t) \equiv (\psi_{1,\mathbf{p}}, \psi_{2,\mathbf{p}}, \cdots, \psi_{q,\mathbf{p}})^T$, $I_q$ is a $q\times q$ identity matrix, and $\hat{E}(\mathbf{p})$ is the diagonal matrix whose $n$-$n$ component is $E_n (\mathbf{p})$. The Berry connection matrix $\overleftrightarrow{\boldsymbol{\mathcal{A}}}(\mathbf{p})$ is a $q\times q$ matrix whose $n$-$n^\prime$ component is $\boldsymbol{\mathcal{A}}_{n,n^\prime} (\mathbf{p})$.
Therefore, the time evolution of the momentum-space wave function $\hat{\psi}_\mathbf{p}$ follows the momentum-space Hamiltonian
\begin{align}
	\overleftrightarrow{\mathcal{H}}^\mathrm{M}
	\equiv
	\hat{E}(\mathbf{p})
	+
	\frac{\kappa}{2}
	\left( I_{q} i\nabla_\mathbf{p} + \overleftrightarrow{\boldsymbol{\mathcal{A}}} (\mathbf{p})\right)^2. \label{hammomentum}
\end{align}
The mapping from the original Hamiltonian (\ref{hamoriginal}) to the momentum-space Hamiltonian (\ref{hammomentum}) is exact with no approximation involved so far.

Now, we assume that the trapping potential and the temperature are small enough that only the lowest band is populated. Then, the dynamics of the momentum-space wave function takes place only within the lowest band. Thus, $\psi_{1,\mathbf{p}}$ is the only nonzero component of the wave function. Its time evolution obeys $i\partial_t \psi_{1,\mathbf{p}}(t) = \mathcal{H}^\mathrm{M} \psi_{1,\mathbf{p}}(t)$, with an effective Hamiltonian~\cite{Berceanu:2015, Claassen:2015}
\begin{align}
	\mathcal{H}^\mathrm{M}
	\approx
	E_1 (\mathbf{p})
	+
	\frac{\kappa}{2}\sum_{n \neq 1} \left| \boldsymbol{\mathcal{A}}_{1,n}(\mathbf{p}) \right|^2
	+
	\frac{\kappa}{2}
	\left[ i\nabla_\mathbf{p} + \boldsymbol{\mathcal{A}}_{1,1}(\mathbf{p})\right]^2. \label{hamsingle}
\end{align}
One may notice a clear analogy between this momentum-space Hamiltonian and the Hamiltonian of a charged particle in an electromagnetic vector potential. We effectively have, in momentum space, a charged particle with mass $\kappa^{-1}$ subject to a vector potential $\boldsymbol{\mathcal{A}}_{1,1}(\mathbf{p})$ and a scalar potential $E_{1} (\mathbf{p}) + (\kappa /2 )\sum_{n\neq 1}|\boldsymbol{\mathcal{A}}_{1,n}(\mathbf{p})|^2$. The momentum-space magnetic field is given by the Berry curvature in momentum space $\Omega (\mathbf{p}) \equiv \nabla_\mathbf{p} \times \boldsymbol{\mathcal{A}}_{1,1}(\mathbf{p})$. It is this analogy with the charged particle in a magnetic field which enables us to explore the physics of quantum Hall effect in momentum space. For simplicity, from now on, we write $\boldsymbol{\mathcal{A}}_{1,1}$ as $\boldsymbol{\mathcal{A}} = (\mathcal{A}_x, \mathcal{A}_y)$.
The total amount of flux going through momentum space is given by the integral of the Berry curvature over the Brillouin zone, i.e., the usual topological Chern number in momentum space. To explore analog magnetic effects, such as quantum Hall effects, in momentum space, we therefore need a model with a non-zero Chern number in momentum space, and then to add an external harmonic trap to this model.

We now allow the trap center $\mathbf{r}_0$ to be at a general position, and show that the trap center $\mathbf{r}_0$ acts as an effective magnetic flux inserted through the holes of the torus of the Brillouin zone (Fig.~\ref{insertingflux}).
The term $\kappa (i\nabla_\mathbf{p} + \boldsymbol{\mathcal{A}})^2/2$ in (\ref{hamsingle}) comes from the harmonic trapping potential $\kappa r^2/2$; $\mathbf{r} \equiv (x,y)$ is replaced by its momentum-space representation $i\nabla_\mathbf{p} + \boldsymbol{\mathcal{A}}$. Then, when the trap center is shifted, the trapping potential has the form $\kappa (\mathbf{r} - \mathbf{r}_0)^2/2$, and the momentum-space Hamiltonian becomes~\cite{Ozawa:2015}
\begin{align}
	\mathcal{H}^\mathrm{M}
	\approx
	E_1 (\mathbf{p})
	+
	\frac{\kappa}{2}\sum_{n \neq 1} \left| \boldsymbol{\mathcal{A}}_{1,n} \right|^2
	+
	\frac{\kappa}{2}
	\left( i\nabla_\mathbf{p} + \boldsymbol{\mathcal{A}}_0 - \mathbf{r}_0 \right)^2, \label{hamshift}
\end{align}
where $\boldsymbol{\mathcal{A}}_0$ is the Berry connection when $\mathbf{r}_0 = 0$.
When the trap center is at a generic position, we can identify a shifted Berry connection $\boldsymbol{\mathcal{A}} \equiv \boldsymbol{\mathcal{A}}_0 - \mathbf{r}_0$, which immediately shows how shifting the trap center corresponds to adding a uniform component in the Berry connection.
As discussed in Sec.~\ref{reviewniu} for the real-space torus, this uniform component can be seen either as a flux through the Brillouin zone, or, equivalently, as a twisted boundary condition in the Brillouin zone. As a result, by changing the position $\mathbf{r}_0$ of the trap center, one can control the amount of the effective flux threading the holes of the Brillouin zone torus.

In Fig.~\ref{insertingflux}, we plot the directions of the magnetic fluxes inserted in the momentum-space torus corresponding to the shift of the trap center in both directions. The amount of flux inserted in the Brillouin zone corresponding to the trap center at $x_0$ is $\Phi_x = \mathcal{L}_x x_0$, where $\mathcal{L}_x$ is the length of the Brillouin zone in the $p_x$ direction [Fig.~\ref{insertingflux}(b)]. Insertion of one magnetic flux quantum corresponds to the trap center position of $2\pi / \mathcal{L}_x$, which in turn corresponds to the length of the unit vector of the real-space lattice in the $x$ direction. Therefore, moving the trap center by one unit vector in real space corresponds to inserting one magnetic flux quantum through a hole of the momentum-space torus. Moving the trap center in the $y$ direction in real space corresponds to inserting a magnetic flux of $\Phi_y = \mathcal{L}_y y_0$ in momentum space in the other direction [Fig.~\ref{insertingflux}(c)]. In an alternative but fully equivalent picture, shifting the trap center from $(0,0)$ to $(x_0, y_0)$ can also be viewed as fixing the Berry connection to $\boldsymbol{\mathcal{A}}_0$ and changing the boundary condition at the edges of the Brillouin zone by factors of $e^{i\mathcal{L}_x x_0}$ and $e^{i\mathcal{L}_y y_0}$ in the $p_x$ and $p_y$ directions, respectively.

Even though we have so far discussed the mapping at the level of the single-particle physics only, it is straightforward to extend it to a many-body system including inter-particle interactions. The momentum-space many-body Hamiltonian with $N$ particles is then
\begin{align}
	\mathcal{H}^\mathrm{M}
	=
	\sum_{a=1}^N
	&\left\{
	E_1 (\mathbf{p}_a)
	+
	\frac{\kappa}{2}\sum_{n \neq 1} \left| \boldsymbol{\mathcal{A}}_{1,n} (\mathbf{p}_a) \right|^2
	\right.
	\notag \\
	&\left.
	+
	\frac{\kappa}{2}
	\left( i\nabla_{\mathbf{p}_a} + \boldsymbol{\mathcal{A}}_0 (\mathbf{p}_a) - \mathbf{r}_0 \right)^2
	\right\}
	+
	\mathcal{H}_\mathrm{int}^\mathrm{M}, \label{hmfull}
\end{align}
where particles are labeled by the index $a$. It is however crucial to note that a local and short-ranged real space interaction translates into a non-local and long-ranged momentum space interparticle interaction $\mathcal{H}_\mathrm{int}^\mathrm{M}$. An example of such a momentum-space interaction is given in~\cite{Ozawa:2015} for the case of the harmonically-trapped Harper-Hofstadter model.

\section{Quantum Hall effect in momentum space}
\label{msqhe}

We are finally in a position to discuss the momentum-space analog of the quantum Hall effect. This effect can be observed by looking at the momentum-space current response as an analog magnetic flux is inserted through a hole of the torus-shaped Brillouin zone. We consider a generic many-body system, described by the momentum-space Hamiltonian~(\ref{hmfull}). Since the insertion of a flux corresponds to a shift in the trap center, we consider adiabatically moving the trap center along the $x$ direction according to a generic smooth function $x_0 (t)$ (a fully analogous analysis holds for the motion of the trap center in the $y$ direction). The current operator can be defined in momentum space in analogy to the real-space current operator~(\ref{totalcurrent}) as~\cite{Price:2015}
\begin{align}
	\tilde{\mathbf{j}}
	=
	\sum_{a=1}^N
	\kappa \left[-i\nabla_{\mathbf{p}_a} - \boldsymbol{\mathcal{A}}(\mathbf{p}_a) \right]
	=
	\nabla_{\mathbf{r}_0}\mathcal{H}^\mathrm{M}, \label{jdef}
\end{align}
where $\nabla_{\mathbf{r}_0}$ is the derivative with respect to $\mathbf{r}_0$.

We assume that we start from the many-body ground state $|\psi_1 \rangle$ and adiabatically move the trap center in the $x$ direction. The momentum-space current in the $p_y$ direction is given by
\begin{align}
	\langle \tilde{j}_y \rangle
	=
	\langle \psi (t) | \partial_{y_0}\mathcal{H}^\mathrm{M}(t) | \psi (t) \rangle, \label{jymean}
\end{align}
where $|\psi (t) \rangle$ is the many-body state at time $t$.
The state $|\psi (t) \rangle$ can be expressed in terms of instantaneous many-body eigenstates $|\psi_\nu(t)\rangle$ of the momentum-space Hamiltonian, labeled by an index $\nu$, which satisfy $\mathcal{H}^\mathrm{M}(t) |\psi_\nu (t)\rangle = \mathcal{E}_\nu (t) |\psi_\nu (t)\rangle$. Here, $\mathcal{E}_\nu (t)$ is the $\nu$-th instantaneous eigenvalue of $\mathcal{H}^\mathrm{M} (t)$ at time $t$, which is also the $\nu$-th eigenvalue of the original real-space Hamiltonian $\mathcal{H}$ at time $t$. Note the difference between $E_n$ and $\mathcal{E}_\nu$; the former is the band dispersion of the periodic part of the Hamiltonian $\mathcal{H}_0$, whereas the latter is the energy of the full Hamiltonian $\mathcal{H}$. Then, up to the first order in time-dependent perturbation theory, we have~\cite{Thouless:1983}
\begin{align}
	|\psi (t)\rangle
	\approx
	|\psi_1 (t)\rangle
	-i\sum_{\mu\neq 1}
	\frac{\langle \psi_{\mu} (t) | \partial_t |\psi_1 (t) \rangle}{\mathcal{E}_1 - \mathcal{E}_{\mu}} |\psi_{\mu} (t)\rangle.
\end{align}
Inserting this expression into (\ref{jymean}), we obtain
\begin{align}
	\langle \tilde{j}_y \rangle
	&=
	\frac{\partial \mathcal{E}_1}{\partial y_0}
	-i
	\left(
	\left\langle \left.\frac{\partial \psi_1(t)}{\partial y_0}\right| \frac{\partial \psi_1 (t)}{\partial t}\right\rangle
	-
	\mathrm{c.c.}
	\right)
	\notag \\
	&=
	\frac{\partial \mathcal{E}_1}{\partial y_0}
	+\frac{\partial x_0}{\partial t}i
	\left(
	\left\langle \left.\frac{\partial \psi_1(t)}{\partial x_0}\right| \frac{\partial \psi_1 (t)}{\partial y_0}\right\rangle
	-
	\mathrm{c.c.}
	\right)
	\notag \\
	&\equiv
	\frac{\partial \mathcal{E}_1}{\partial y_0}
	+\frac{\partial x_0}{\partial t} \tilde{\Omega}(\mathbf{r}_0), \label{hallcurrent}
\end{align}
where $\tilde{\Omega}(\mathbf{r}_0)$ is the Berry curvature in $\mathbf{r}_0$ space.

The connection between $\langle \tilde{j}_y \rangle$ and the quantum Hall effect may be more transparent when one writes $\partial_t x_0 = \partial_t \Phi_x / \mathcal{L}_x$. The term $\partial_t \Phi_x$ acts as an ``emf" in momentum space coming from the Faraday's law in momentum space, so $\partial_t x_0$ is nothing but the induced artificial electric field in momentum space. Thus, the Hall conductivity in momentum space is given by
\begin{align}
	\tilde{\sigma}_{yx}
	=
	-\tilde{\sigma}_{xy}
	=
	\tilde{\Omega}(\mathbf{r}_0),
\end{align}
which is exactly the analog of the real-space Hall conductivity~(\ref{rsxy}). This expression shows that the Niu-Thouless-Wu formulation of the quantum Hall effect applies also to momentum space. Note that generally speaking there can be a momentum-space current even in the absence of the artificial electric field due to the first term of~(\ref{hallcurrent}), which gives the persistent current in momentum space~\cite{Price:2015}. Such a persistent current can also exist in the real-space quantum Hall effect; an analogous expression for the real-space Hall current including the persistent current can be found, for example, in~\cite{Avron:1988}.
As noted in the Introduction, our quantum Hall effect in momentum space can be viewed as the dynamical quantum Hall effect where the dynamical parameter is the trap center position $\mathbf{r}_0$~\cite{Gritsev:2012}.

The momentum-space current derived above is an experimentally observable quantity. In fact, since $\mathbf{r} = i\nabla_\mathbf{p} + \boldsymbol{\mathcal{A}}_0$ from the minimal coupling in momentum space, we have from (\ref{jdef}) that
\begin{align}
	\tilde{\mathbf{j}} = -\kappa \sum_{a=1}^N \left( \mathbf{r}_a - \mathbf{r}_0 \right). \label{jobservable}
\end{align}
Thus, the momentum-space current is just the mean position of the wave function measured from the trap center times a constant factor $(-\kappa)$.
In this paper, we refer to the expression (\ref{hallcurrent}) as the \textit{geometrical} expression of the momentum-space current, and the expression (\ref{jobservable}) as the \textit{observable} expression.
In the next section, we apply the general theory developed in this section to a particular model, a trapped Harper-Hofstadter model, and numerically verify that the geometrical and observable expressions for the momentum-space current indeed agree.

In order to experimentally observe the ``quantized" nature of the momentum-space Hall current, it is convenient to consider an analog of the net adiabatic charge transport after the trap center is moved by one unit cell in real space from $x_0 (0) = 0$ to $x_0 (T) = 2\pi/\mathcal{L}_x$. The charge transport in momentum space is given by
\begin{align}
	P_y
	\equiv
	\int_0^T dt \langle \tilde{j}_y \rangle
	=
	\int_0^T dt \frac{\partial \mathcal{E}_1}{\partial y_0}
	+
	\int_0^{\frac{2\pi}{\mathcal{L}_x}}dx_0 \tilde{\Omega}(\mathbf{r}_0). \label{pump}
\end{align}
The first term, coming from the momentum-space persistent current, depends on the particular way $x_0 (t)$ is realized, while the second term depends only on the geometrical properties of the momentum-space model and is independent of the particular time dependence of $x_0 (t)$.
By taking the average of the measured $P_y$ for different values of $y_0$, the persistent current term vanishes and one obtains
\begin{align}
	\bar{P}_y
	&\equiv
	\frac{\mathcal{L}_y}{2\pi}\int_0^{\frac{2\pi}{\mathcal{L}_y}}dy_0 P_y
	\notag \\
	&=
	\frac{\mathcal{L}_y}{2\pi}\int_0^{\frac{2\pi}{\mathcal{L}_y}}dy_0
	\int_0^{\frac{2\pi}{\mathcal{L}_x}}dx_0 \tilde{\Omega}(\mathbf{r}_0)
	\notag \\
	&=
	\mathcal{L}_y \tilde{\mathcal{C}}, \label{quantizedpump}
\end{align}
where $\tilde{\mathcal{C}}$ is the Chern number in $\mathbf{r}_0$ space.
Thus, $\bar{P}_y/ \mathcal{L}_y$ is quantized to an integer value.

The observable expression for the charge transport in momentum space is
\begin{align}
	P_y
	=
	-\kappa \sum_{a=1}^N \int_0^T dt \left( \langle y_a (t) \rangle - y_0 \right), \label{pumpobservable}
\end{align}
which can be estimated by measuring the mean position of the wave function as the trap center moves.
Experimentally, the quantization of $\bar{P}_y / \mathcal{L}_y$ to $\tilde{\mathcal{C}}$ can be checked by repeating the measurement of (\ref{pumpobservable}) for different values of $y_0$.

This averaging procedure is to be contrasted with the real-space quantum Hall effect where multiple experiments with different twisted boundary conditions are not needed to obtain the quantized Hall conductance in a large system approaching the thermodynamic limit~\cite{Niu:1985, Niu:1987}. This is because, in the thermodynamic limit, the Berry curvature in the parameter space of the twist boundary condition becomes flat, so only one value of the boundary condition suffices to make the charge transport quantized. In contrast, the thermodynamic limit cannot be taken in the momentum-space quantum Hall effect, as the size of the Brillouin zone is not an extensive quantity, so, generally speaking, the averaging as done in (\ref{quantizedpump}) is necessary to obtain a quantized value. However, as we discuss later, in suitable limits where the Berry curvature $\tilde{\Omega} (\mathbf{r}_0)$ becomes flat and the persistent current vanishes, the quantization
\begin{align}
	P_y/\mathcal{L}_y \approx \tilde{\mathcal{C}} \label{quantizedpumpapprox}
\end{align}
holds even when $P_y$ is calculated for one particular value of $y_0$.

\section{Trapped Harper-Hofstadter model}
\label{thhm}

We now apply the theory developed in the previous sections to a particular model, namely, the Harper-Hofstadter model: a two-dimensional tight-binding model with a uniform perpendicular magnetic field. The Hamiltonian of the Harper-Hofstadter model on a square lattice with a magnetic flux $2\pi \alpha$ piercing through a plaquette is
\begin{align}
	\mathcal{H}_0
	=
	-J\sum_{x,y} \left( \hat{a}_{x+1,y}^\dagger \hat{a}_{x,y} + e^{i2\pi \alpha x}\hat{a}_{x,y+1}^\dagger \hat{a}_{x,y} + \mathrm{H.c.} \right), \label{hamhh}
\end{align}
where $\hat{a}_{x,y}$ is the annihilation operator of a particle at a site $(x,y)$, and $J > 0$ is the hopping amplitude. We assume the lattice spacing to be $1$, and that $x$ and $y$ are both integers.
Writing $\alpha = p/q$, where $p$ and $q$ are coprime integers, the Harper-Hofstadter model has $q$ bands, which are all topologically nontrivial.
Our gauge choice in (\ref{hamhh}) corresponds to the Landau gauge, and the Hamiltonian has the translational invariance as $x \to x + q$ and $y \to y+1$.
This implies that the most natural choice of the Brillouin zone has lengths $\mathcal{L}_x = 2\pi/q$ and $\mathcal{L}_y = 2\pi$.

We consider the Harper-Hofstadter model coupled to a harmonic trapping potential, so that the total single-particle Hamiltonian is
\begin{align}
	\mathcal{H} =
	\mathcal{H}_0 + \frac{1}{2}\kappa \sum_{x,y} \{(x-x_0)^2 + (y-y_0)^2\} a_{x,y}^\dagger a_{x,y}.
	\label{ourham}
\end{align}
where, as before, $\kappa$ is the strength of the harmonic trapping potential. We start our discussion by focusing in this section on the non-interacting single-particle case and we postpone the discussion of the weakly interacting case to Sec.~\ref{sec:interaction}. The momentum-space effective model of (\ref{ourham}) and its band structure and ground states have been discussed at length in~\cite{Scaffidi:2014,Price:2014,Ozawa:2015}. We now explore the momentum-space quantum Hall effect of this model. 

Note that in this model, the single-particle ground state is nondegenerate provided one works with a Harper-Hofstadter band with a momentum-space Chern number $|\mathcal{C}| = 1$; in this case, the nonzero gap to the first excited state is set by the trapping potential. In suitable regimes, this gap has the physical meaning of the momentum-space cyclotron energy~\cite{Price:2014}.

\subsection{Momentum-space persistent current}

We first focus on the first term in (\ref{hallcurrent}), which is the persistent current in momentum space.
Comparing geometrical (\ref{hallcurrent}) and observable (\ref{jobservable}) expressions for the momentum-space current, when the trap center is not moving ($\partial_t x_0 = 0$) we have
\begin{align}
	-\frac{\langle \tilde{j}_y \rangle}{\kappa}
	=
	\langle y \rangle - y_0
	=
	-\frac{1}{\kappa}
	\frac{\partial \mathcal{E}_0}{\partial y_0}. \label{persistenteq}
\end{align}
This relation implies that the position of the trap center $y_0$ and the center-of-mass position of the wave function $\langle y \rangle$ are generally different. This difference has an interpretation as the persistent current in momentum space, and it is related to the analog of the group velocity $\partial_{y_0} \mathcal{E}_0$ in $\mathbf{r}_0$ space.

\begin{figure}[htbp]
\begin{center}
\includegraphics[width=7.0cm]{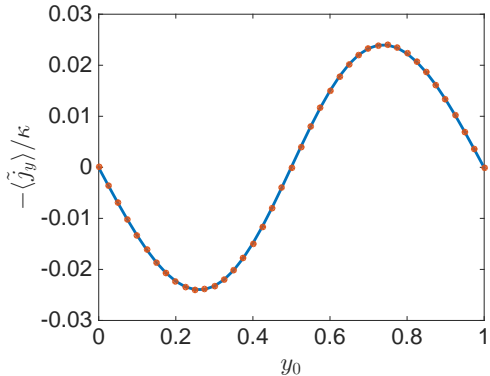}
\caption{Persistent current contribution to the Hall current $-\langle \tilde{j}_y\rangle/\kappa$ as a function of $y_0$ for $x_0 = 0$,  $\alpha = 1/4$, and $\kappa = 0.01J$. The solid line is the geometrical right-hand side of (\ref{persistenteq}), and dots are the observable middle expression of (\ref{persistenteq}) calculated from the non-interacting ground state.}
\label{pers}
\end{center}
\end{figure}

We now numerically confirm the relation~(\ref{persistenteq}).
In Fig.~\ref{pers}, we plot the middle and right expressions of (\ref{persistenteq}) for $\alpha = 1/4$ and $\kappa = 0.01J$. We estimate the middle (observable) expression in (\ref{persistenteq}) by numerically obtaining the ground-state wave function of (\ref{ourham}) and calculating $\langle y \rangle - y_0$. The right (geometrical) expression in (\ref{persistenteq}) is instead estimated through the numerically obtained ground-state energy of (\ref{ourham}). We see a perfect agreement between these two different methods to calculate the persistent current in momentum space.

\subsection{Geometrical contribution to the momentum-space current}

\begin{figure*}[htbp]
\begin{center}
\subfigure[$x_0$ (solid) and $\partial_t x_0$ (dashed) as a function of $t$]{
\includegraphics[width=5.5cm]{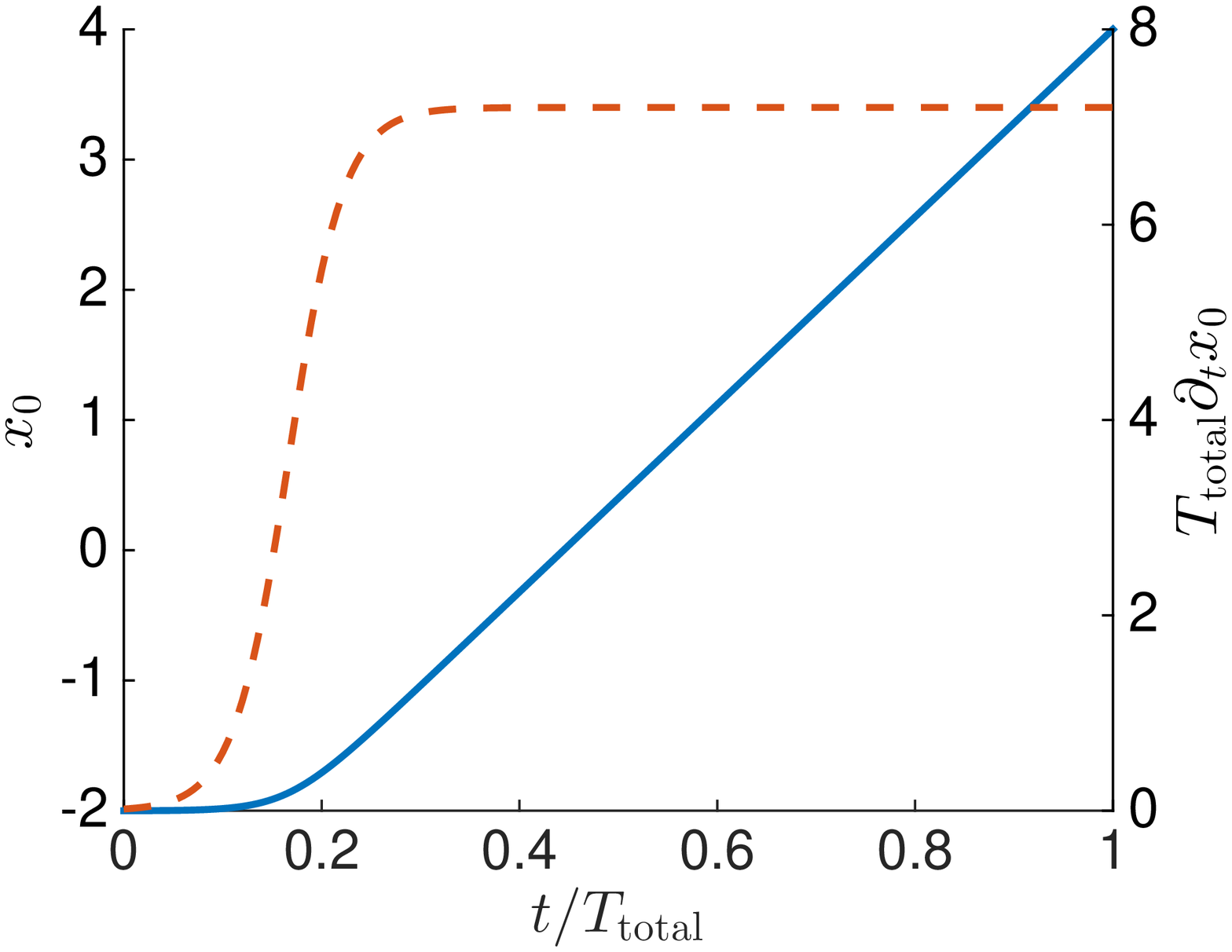}}
\subfigure[Comparison of $\langle y \rangle$ and $-v \tilde{\Omega}(\mathbf{r}_0)/\kappa$ for $T_{\mathrm{total}} = 10^5/J$]{
\includegraphics[width=5.5cm]{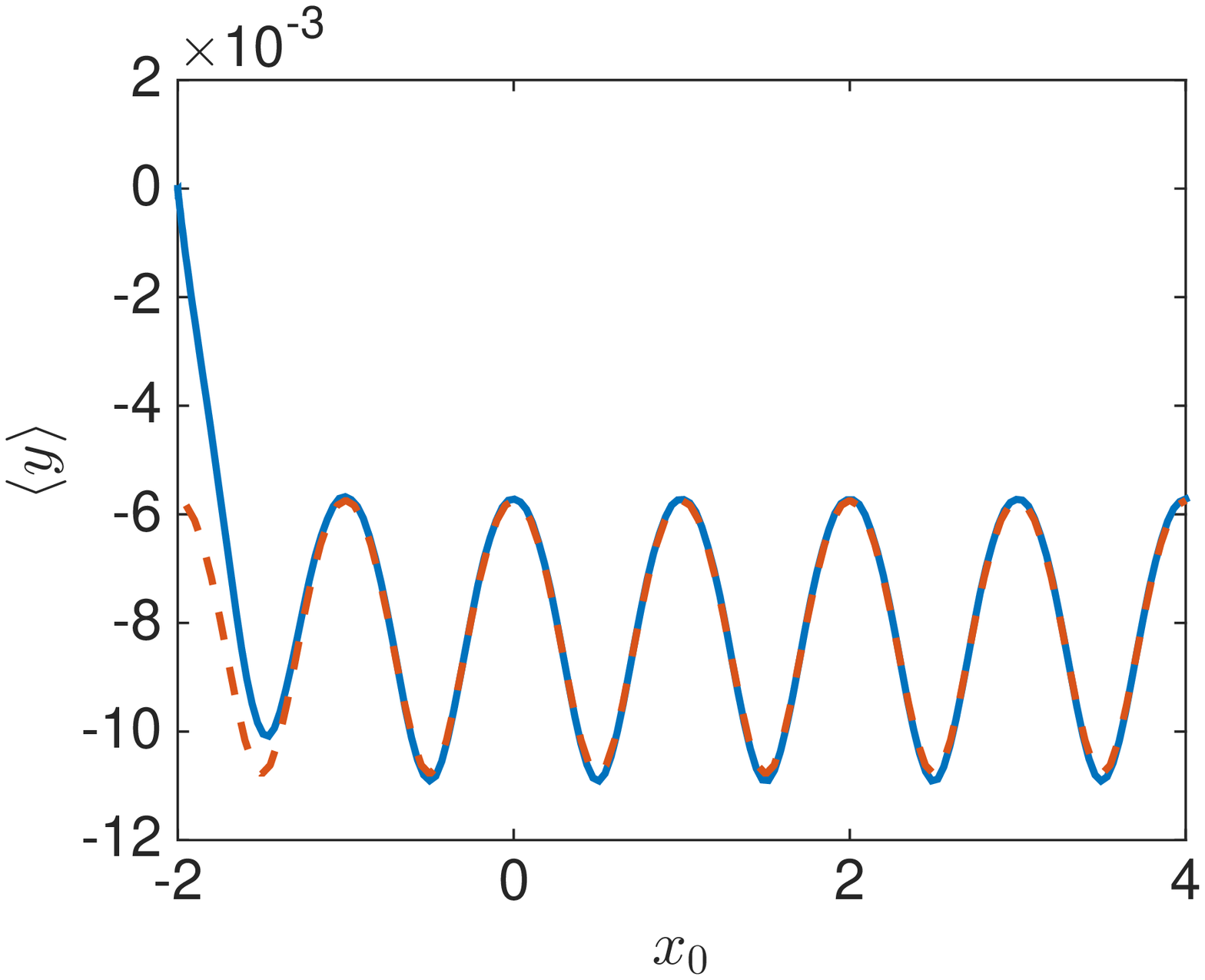}}
\subfigure[Comparison of $\langle y \rangle$ and $-v \tilde{\Omega}(\mathbf{r}_0)/\kappa$ for $T_{\mathrm{total}} = 5\times 10^4/J$]{
\includegraphics[width=5.5cm]{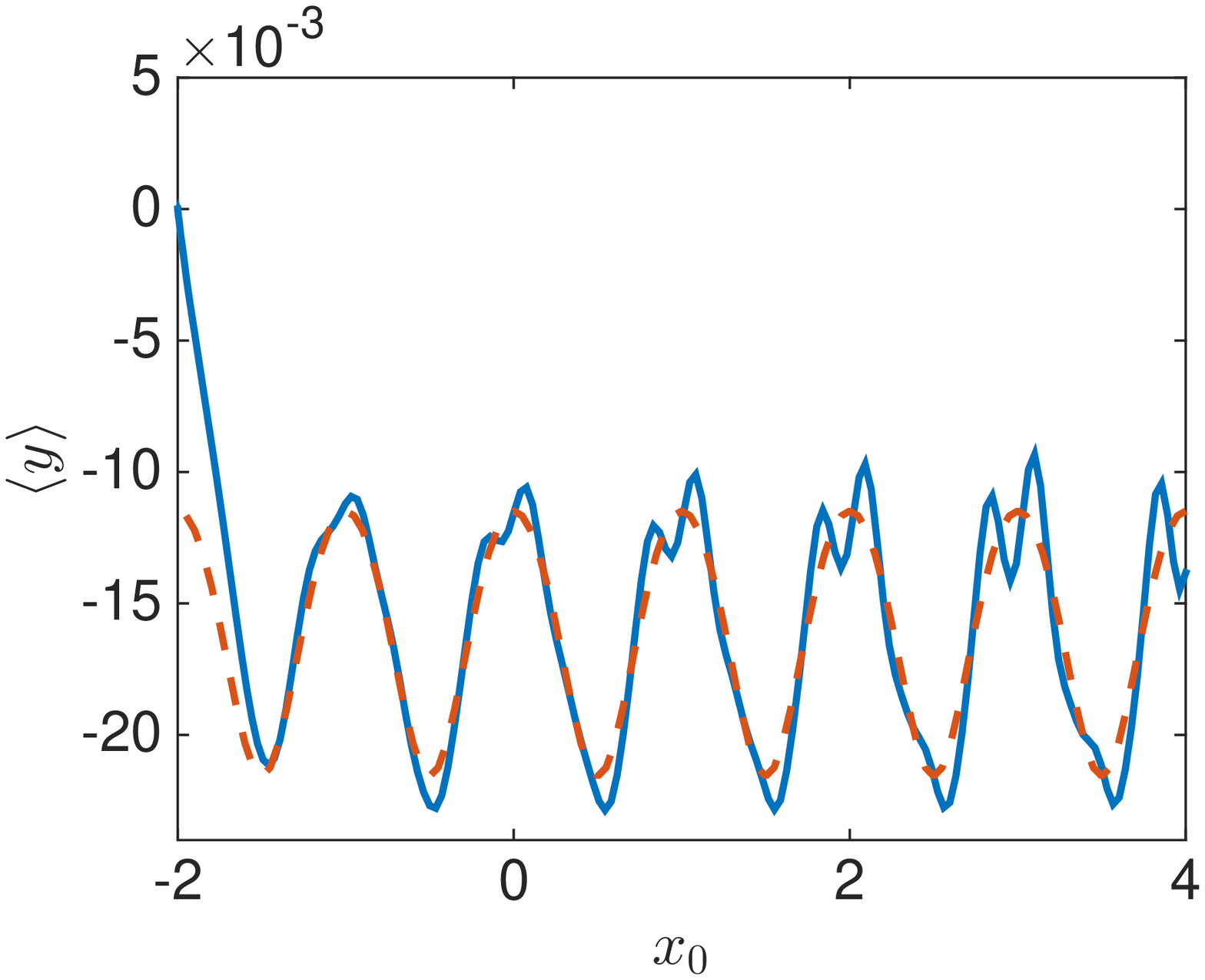}}
\caption{Simulation of the momentum-space Hall conductivity for $\alpha = 1/4$ and $\kappa = 0.01J$. (a) The position of the center of the trap $x_0$ (solid line and the left axis) and the velocity of the trap motion $\partial_t x_0$ (dashed line and the right axis) are plotted as a function of time in units of the total sweeping time $T_{\mathrm{total}}$. (b), (c) The mean position of the wave function $\langle y \rangle$ as a function of the center of the trap $x_0$. The solid line is the numerically obtained value of the observable $\langle y\rangle$ from the real-time simulation starting from the ground state, whereas the dashed line is the geometrical prediction $-v \tilde{\Omega}(\mathbf{r}_0)/\kappa$. The total sweep time is $T_{\mathrm{total}} = 10^5/J$ for (b) and $T_{\mathrm{total}} = 5\times 10^4/J$ for (c). As predicted by Eq.~(\ref{mshalleq}), the two lines agree well for (b) in the region where the velocity is almost constant. The deviation of the two lines seen in (c) is due to the faster sweep causing the transition to excited states.}
\label{hall}
\end{center}
\end{figure*}

Now, we move on to understand the second term in (\ref{hallcurrent}), which accounts for the momentum-space Hall conductivity. It is numerically (and experimentally) easiest to consider the case when the persistent current term is absent, for example, when $\mathbf{r}_0$ is chosen to be along the high-symmetry line at $y_0 = 0$ (see Fig.~\ref{pers}). Then, comparing (\ref{hallcurrent}) and (\ref{jobservable}), one obtains
\begin{align}
	-\frac{\langle \tilde{j}_y \rangle}{\kappa} = \langle y \rangle = -\frac{1}{\kappa} \frac{d x_0 (t)}{dt} \tilde{\Omega}(\mathbf{r}_0).
\end{align}
In particular, when the trap center is moving with a constant velocity $x_0 (t) = vt$, one has the following relation
\begin{align}
	\langle y \rangle = -\frac{v}{\kappa} \tilde{\Omega}(\mathbf{r}_0), \label{mshalleq}
\end{align}
which shows how the mean displacement of the wavefunction $\langle y \rangle$ is directly proportional to the Berry curvature in $\mathbf{r}_0$ space.
This is the analog of the anomalous Hall effect in momentum space.

To quantitatively check this relation, we perform the real-time simulation of the non-interacting Schr\"odinger equation starting from its ground state for a trap center at $\mathbf{r}_0=0$ and then slowly moving the trap center along the $x$ axis. For $\alpha = 1/4$ and $\kappa = 0.01J$, we plot the result of the numerical simulation in Fig.~\ref{hall}.
We change the velocity $\partial_t x_0 \propto \tanh t$ starting from $x_0 = -2$ so that in the interval $0 < x_0 < 4$ the velocity is almost constant. Both $x_0(t)$ and $\partial_t x_0(t)$ are plotted in Fig.~\ref{hall}(a).
The motion of the trap center should be slow enough not to cause any Landau-Zener transition to excited states; in particular, the maximum speed of the trap center $\partial_t x_0(t)$ should be much slower than the minimum gap between the lowest-energy state and the first excited state, which in the current case is $\sim 2.4 \times 10^{-3}J$.

In Figs.~\ref{hall} (b) and (c), the comparisons between the left- and the right- hand sides of (\ref{mshalleq}) are plotted for two different velocities of the trap center sweep.
The (observable) left-hand side of (\ref{mshalleq}) is calculated from the real-time simulation of the wave function according to the full Hamiltonian~(\ref{ourham}). The (geometrical) right-hand side of (\ref{mshalleq}) is instead numerically estimated by calculating the $\mathrm{r}_0$-space Berry curvature through~\cite{Bernevig:Book}
\begin{align}
	\tilde{\Omega}(\mathbf{r}_0)
	=
	-2\, \mathrm{Im}
	\sum_{n\neq 1}
	\frac{\langle \psi_1 | \partial_{x_0} \mathcal{H} | \psi_n \rangle \langle \psi_n | \partial_{y_0} \mathcal{H} | \psi_1 \rangle}
	{(\mathcal{E}_n - \mathcal{E}_1 )^2}. \label{berrynum}
\end{align}
Note that both $|\psi_n\rangle$ and $\mathcal{E}_n$ are functions of $\mathbf{r}_0$.
In Fig.~\ref{hall}(b), we take the total sweep time to be $T_{\mathrm{total}} = 10^5/J$, which corresponds to $v \approx 7.2\times 10^{-5} J$ in the constant velocity region.
In Fig.~\ref{hall}(c), the sweep is twice faster than that in Fig.~\ref{hall}(b), namely, $T_{\mathrm{total}} = 5\times 10^4/J$ and $v \approx 1.44\times 10^{-4} J$ in the constant velocity region.
In Fig.~\ref{hall}(b), perfect agreement is found throughout the whole region where the velocity of the trap center is almost constant, whereas the deviation visible in Fig.~\ref{hall}(c) is due to the faster motion of the trap center which makes the dynamics less adiabatic.

\subsection{Numerical verification of the momentum-space quantum Hall effect}

We now numerically verify the momentum-space quantum Hall effect.
As we mentioned above, one can in principle choose different values of $y_0$, and the average charge transport should be quantized as in (\ref{quantizedpump}). Alternatively, the charge transport even for one value of $y_0$ can approach a quantized value as in (\ref{quantizedpumpapprox}) when the Berry curvature in $\mathbf{r}_0$ space becomes flat. As mentioned in the original Niu-Thouless-Wu paper~\cite{Niu:1985}, this is the case when the underlying potential is flat and the interparticle interaction is negligible. In our momentum-space setup, the underlying potential becomes flat compared to the band gap and/or the kinetic energy in the limit of large $\kappa$ or small $\alpha$. As in these limits one can approximate the momentum-space magnetic field as being uniform across the whole magnetic Brillouin zone, one expects that the quantization of the charge transport is observable without taking an average. This statement will be numerically confirmed in the following.

We consider either $y_0 = 0$ or $0.5$, in which case the persistent current contribution to the momentum-space current vanishes. We sweep the trap center from $x_0 = 0$ to 1 according to a sinusoidal time dependence $x_0 (t) = [1-\cos (\pi t/T)]/2$.
Note that, unlike in the previous subsection where we probed the geometrically local property $\tilde{\Omega}(\mathbf{r}_0)$ by moving the trap center at a constant speed, the charge transport (\ref{pump}) does not depend on a particular form of $x_0 (t)$ and thus does not require the sweep of the trap center to have a constant speed.
The geometrical and observable expressions of the charge transport satisfy the following relation:
\begin{align}
	\frac{P_y}{\mathcal{L}_y}
	=
	\frac{q}{2\pi}\int_0^{1} dx_0 \tilde{\Omega}(\mathbf{r}_0)
	=
	-\frac{q \kappa}{2\pi}\int_0^T dt [\langle y(t)\rangle - y_0], \label{pyoverly}
\end{align}
which should recover the quantized value of $\tilde{\mathcal{C}}$ in the limit of flat $\tilde{\Omega}(\mathbf{r}_0)$.
For the trapped Harper-Hofstadter model with $\alpha = 1/q$, the Chern number $\tilde{\mathcal{C}}$ in $\mathbf{r}_0$ space is always $+1$, as detailed in Appendix~\ref{sec:app}.

In Fig.~\ref{quantizednumerics}, we plot the estimate of (\ref{pyoverly}) for increasing $\kappa$ and decreasing $\alpha$, for both $y_0 = 0$ and $0.5$. In Fig.~\ref{quantizednumerics}(a), we fix $\alpha = 1/4$ and vary $\kappa$ and compare the  middle and the right expressions of (\ref{pyoverly}). The (geometrical) middle expression is numerically estimated from (\ref{berrynum}), while the (observable) right-hand side of (\ref{pyoverly}) is estimated from the real-time simulation of the Schr\"odinger equation starting from its ground state for the trap center at $\mathbf{r}_0=0$. We see that the estimate of (\ref{pyoverly}) approaches the quantized value $+1$ as $\kappa$ becomes larger. The further increase of $\kappa$ beyond the region plotted in Fig.~\ref{quantizednumerics}(a) results in a strong deviation of $P_y / \mathcal{L}_y$ from $1$, because the single-band approximation we employed upon deriving (\ref{hamsingle}) breaks down for a too large trap strength $\kappa$~\cite{Price:2014}. The appreciable remaining deviation that one can see in the plotted region around, say, $\kappa\approx 0.2$ can be attributed to the significant variation of the energy dispersion and the Berry curvature across the magnetic Brillouin zone for the relatively large $\alpha=1/4$ considered here.

An even more favorable regime is illustrated in Fig.~\ref{quantizednumerics}(b), where we plot the middle and the right expressions of (\ref{pyoverly}) for a fixed value of $\kappa = 0.01J$ and varying $\alpha$. In this case, the transported charge quickly converges to an integer value $+1$ as $\alpha \to 0$, which confirms that the quantization is excellently recovered provided one chooses a suitable regime.

\begin{figure}[htbp]
\begin{center}
\subfigure[large $\kappa $ limit]{
\includegraphics[width=6.0cm]{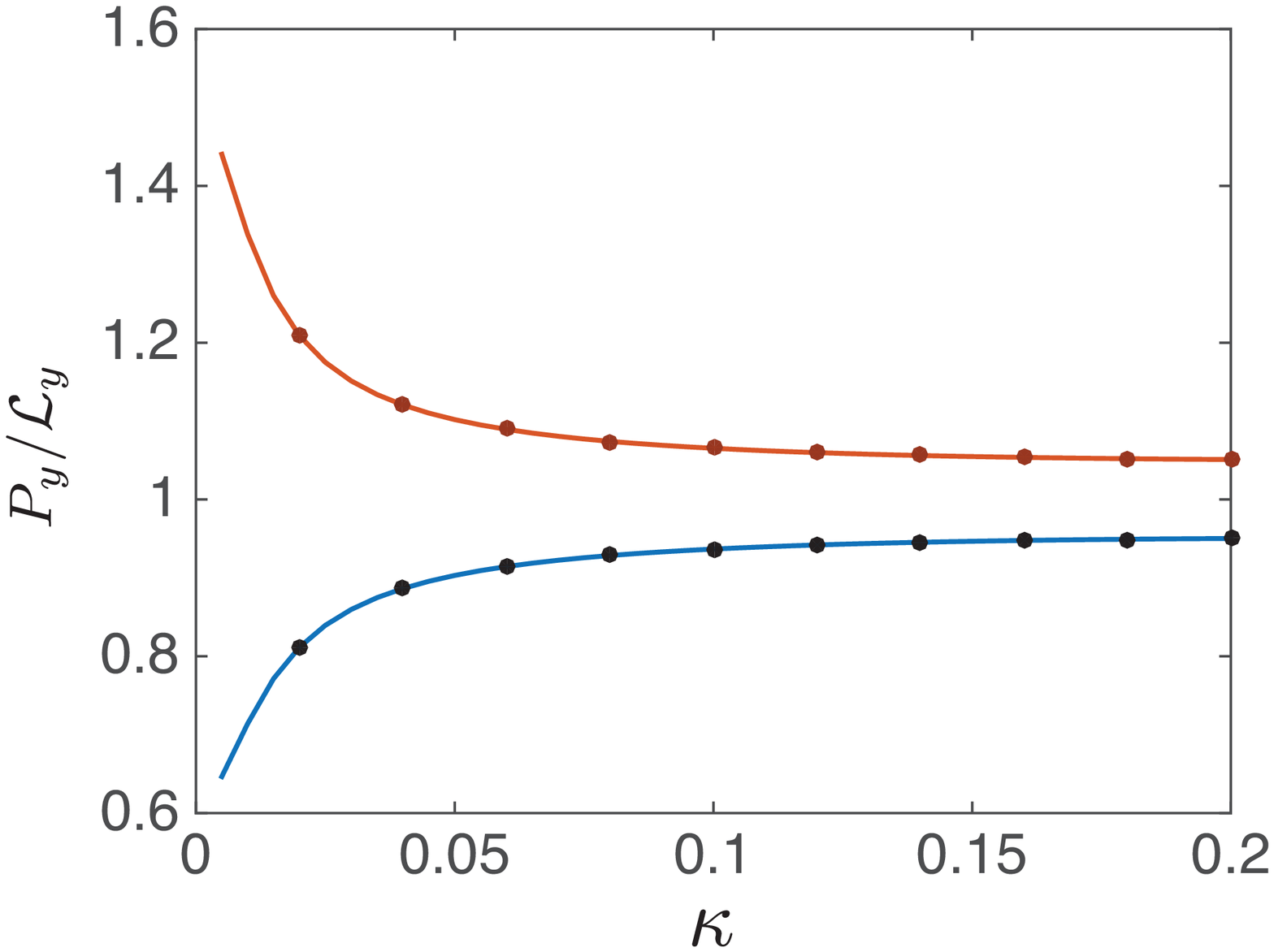}}
\subfigure[small $\alpha$ limit]{
\includegraphics[width=6.0cm]{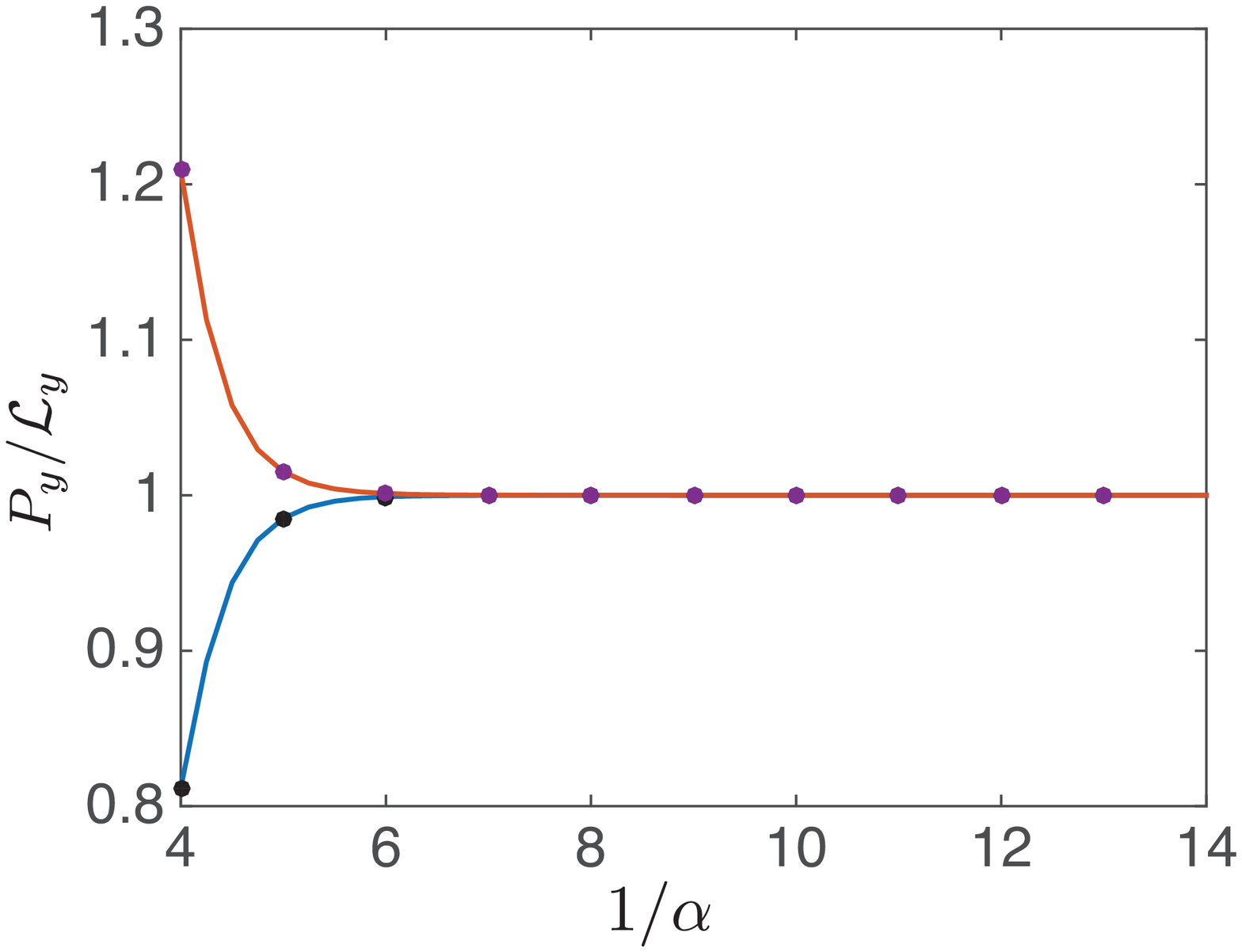}}
\caption{Estimated charge transport $P_y / \mathcal{L}_y$. The lines are the geometrical contribution to the charge transport (the middle expression of (\ref{pyoverly})), and the dots are estimated from the numerical calculation using the observable expression of the charge transport (the right expression of (\ref{pyoverly})). The lower line and dots are for $y_0 = 0$ and the upper line and dots are for $y_0 = 0.5$. (a) $\alpha = 1/4$ is fixed and $\kappa$ is varied. (b) $\kappa = 0.01J$ is fixed and $\alpha$ is varied.}
\label{quantizednumerics}
\end{center}
\end{figure}

\section{Interacting system}
\label{sec:interaction}

In the previous section, we have studied the momentum-space integer quantum Hall effect for non-interacting particles. In the presence of a very weak interaction, the ground state for bosons is still unique and gapped and one can expect that the integer quantum Hall effect in momentum space is still observed with no qualitative nor quantitative change.

The situation is far more interesting for stronger interactions, but still in the mean-field regime of dilute Bose gases. For this regime, we have previously found within mean-field theory that the harmonically trapped Harper-Hofstadter model can have degenerate ground states that spontaneously break a rotational symmetry~\cite{Ozawa:2015}. From the general theory of the quantum Hall effect~\cite{Niu:1985, Oshikawa:2006}, we know that a topological degeneracy of the ground state in a toroidal geometry typically implies fractional states. It is therefore natural to wonder if the degeneracy found in~\cite{Ozawa:2015} for a mean-field regime can be somehow related to some kind of fractional states in momentum space and if one can access the fractional quantum Hall effect in momentum space. Although this is unfortunately not the case, as we discuss below, it is interesting to consider why the previously found ground-state degeneracy does not imply a fractional quantum Hall effect in momentum space. While this negative result is, in a way, not surprising as the interaction we are considering is a mean-field one, it does not exclude at all the possibility of obtaining fractional states if one includes stronger beyond-mean-field interactions, as will be the subject of future work.

\begin{figure}[htbp]
\begin{center}
\includegraphics[width=8.0cm]{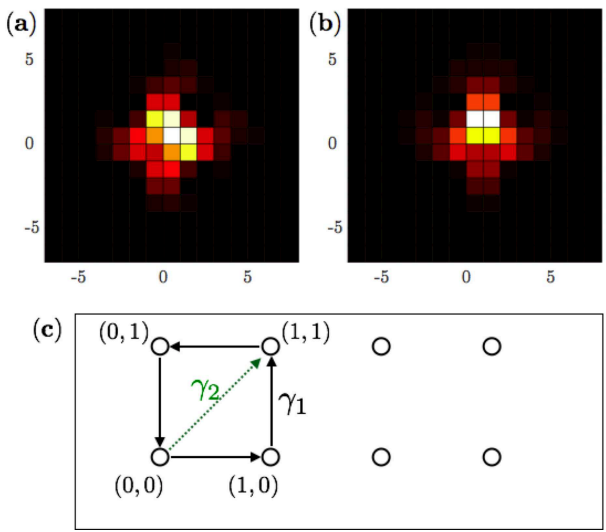}
\caption{Spatial density profile of one of the four degenerate ground states of the trapped Harper-Hofstadter model with $\kappa = 0.02J$ and $UN = 0.5J$ when the trap center is located at (a) $\mathbf{r}_0 = (0,0)$ and (b) $\mathbf{r}_0 = (1/2,1/2)$. (c) Schematic drawing of the real-space lattice displaying the two trajectories $\gamma_1$ (along solid black arrows) and $\gamma_2$ (along a dotted green arrow) discussed in the main text for moving the center of the trap.}
\label{interaction}
\end{center}
\end{figure}

A typical example of a degenerate, symmetry-breaking ground state of the interacting trapped Harper-Hofstadter model when the trap center is at the origin, $\mathbf{r}_0 = (0,0)$, is depicted in Fig.~\ref{interaction}(a). A characteristic feature of this state is that the density is concentrated in the first quadrant. Since the $90^\circ$ rotation modified with an appropriate phase is a symmetry of the system~\cite{Ozawa:2015,Balents:2005,Powell:2010,Powell:2011}, one can construct three other degenerate ground states from the state in Fig.~\ref{interaction}(a) by modified rotations. 

We consider moving the trap center along the square-shaped trajectory $\gamma_1$ described in Fig.~\ref{interaction}(c) along the path $(0,0) \to (1,0) \to (1,1) \to (0,1) \to (0,0)$. Tracking how a ground state evolves as the trap center follows $\gamma_1$, we have found that when the trajectory is completed and the trap center comes backs to the original point $(0,0)$, the ground state is transformed into a different degenerate ground state.

This strange transformation of the ground states implies that there is at least one point within the region enclosed by the trajectory where the adiabatic motion is not well defined. One can see this by the following simple thought experiment. We can imagine making the trajectory $\gamma_1$ gradually smaller, while fixing the initial and the final points at the origin $(0,0)$. If the adiabatic motion is everywhere well defined inside the trajectory of $\gamma_1$, making the trajectory smaller does not change the initial and the final states, and the initial and the final states should always be different degenerate ground states. However, when the trajectory is sufficiently small, the initial and the final states should be the same, which is a contradiction. Therefore, there must be a point within the trajectory of $\gamma_1$ where the adiabatic motion is ill defined. The adiabatic motion can be ill defined, for example, either when the gap to excited states close, or when two ground states merge, or when the adiabatic trajectory reaches saddle points in the energy landscape. When the adiabatic motion is not well defined, the Niu-Thouless-Wu formalism fails and it does not give rise to the quantum Hall effect.

In fact, the ground state is four-fold degenerate only at high-symmetry points such as the origin $\mathbf{r}_0 = (0,0)$. For a general value of $\mathbf{r}_0$, the ground state is uniquely determined. At the middle of a plaquette $\mathbf{r}_0 = (1/2,1/2)$, the system again has the four-fold modified rotational symmetry, and the symmetry-breaking ground state is four-fold degenerate; a typical state is depicted in Fig.~\ref{interaction}(b) and shows that the density is concentrated in the upper half-plane.

The reflection symmetry of the ground state at $\mathbf{r}_0 = (0,0)$ [Fig.~\ref{interaction}(a)] and of the one at $\mathbf{r}_0 = (1/2,1/2)$ [Fig.~\ref{interaction}(b)] are therefore different; the former is symmetric with respect to the reflection along a diagonal line, whereas the latter is symmetric with respect to a horizontal or vertical line. Therefore, as one moves the trap center along the diagonal $(0,0) \to (1,1)$ following the trajectory $\gamma_2$ in Fig.~\ref{interaction}(c), the state depicted in Fig.~\ref{interaction}(a) initially keeps the original reflection symmetry along the diagonal, but eventually needs to break this symmetry around the center $\mathbf{r}_0 = (1/2,1/2)$ spontaneously falling into one of the ground states of Fig.~\ref{interaction}(b) with a different symmetry. It is at this central point $\mathbf{r}_0 = (1/2,1/2)$ where the adiabatic motion is ill defined in this mean-field model.

In more formal terms, the absence of fractional quantum Hall physics in mean-field theory can be traced back to a crucial mathematical difference between the degenerate ground states in standard fractional quantum Hall systems in the presence of strong interactions~\cite{Niu:1985, Oshikawa:2006} and the ones predicted by the nonlinear Schr\"odinger equation with the mean-field theory of ~\cite{Ozawa:2015}. 

In the former theory, the degenerate ground states belong to a linear sub-space so that any linear combination of these is also a ground state and the outcome of transport around a loop in $\mathbf{r}_0$ space is a linear operation within the linear space of ground states; in particular, when the loop is contracted to a point, this linear operation tends to the identity.
On the other hand, in the latter case, the degenerate ground states correspond to discrete points in the phase space, so they can not be continuously deformed one into another as one contracts a loop. This explains why contracting a trajectory giving a non-trivial effect on the wave function like $\gamma_1$ must necessarily hit some singular point.

\section{Discussion}
\label{sec:discussion}

Promising experimental systems in which to observe the quantum Hall effect in momentum space are ultracold atomic gases, where the Harper-Hofstadter model has been recently realized~\cite{Aidelsburger:2013, Miyake:2013, Aidelsburger:2014, Kennedy:2015}, and the harmonic trapping potential required to observe the momentum-space quantum Hall effect is naturally present. The mean position of the wave function, which is proportional to the momentum-space Hall current, is directly measurable through the center-of-mass position of the Bose-Einstein condensate. Furthermore, the trap center can be controlled by applying a linear potential gradient whose strength can be varied in a well-controlled manner. Once the system is cooled into a Bose-Einstein condensate~\cite{Kennedy:2015}, the remaining experimental challenges for the quantitative validation of the momentum-space quantum Hall effect are mostly of quantitative nature and reduce to the precise control of the trap center and the very high accuracy in the measurement of the center-of-mass position.

Another promising platform in which to observe the momentum-space quantum Hall effect is offered by photonics systems, where the Harper-Hofstadter model has been realized in coupled cavity arrays~\cite{Hafezi:2013} and a harmonic trapping potential can be readily applied by varying the resonant frequencies of the cavities in a position-dependent and time-dependent manner. In this case, however, a potential difficulty requiring further exploration comes from the interplay of the adiabatic features discussed in this work with the intrinsically driven-dissipative nature of the photonic system~\cite{Berceanu:2015}.

\section{Conclusion}
\label{sec:conclusion}

In this paper, we have investigated a novel form of the quantum Hall effect that takes place in momentum space. Exploiting the fact that the Brillouin zone naturally has a toroidal structure, we have shown that the Niu-Thouless-Wu formalism of the quantum Hall effect in terms of twisted boundary conditions can be directly realized in momentum space. We have numerically verified the integer quantum Hall effect in momentum space by analyzing the harmonically trapped Harper-Hofstadter model. 

Our results open up new perspectives in the study of the quantum Hall effect by allowing one to experimentally access various results for which the toroidal geometry is crucial. Of particular interest is the investigation of fractional quantum Hall states in momentum space, and of how the topological degeneracy of such states could be experimentally probed. Because of the non-local and long-range nature of interactions in momentum space, fractional states in momentum space cannot be obtained straightforwardly from the analogy to the real-space counterpart. Although it has been anticipated that certain types of strong interactions in the presence of a confining potential and a flat dispersion with nontrivial topology can lead to the fractional states in momentum space~\cite{Claassen:2015}, many questions remain open. 
Therefore, the idea of the momentum-space quantum Hall effect has the potential to become an important tool for the experimental study of exotic fractional states.

\begin{acknowledgements}
The authors thank Leonardo Mazza for interesting discussions. This work was funded by ERC through the QGBE grant, by the EU-FET Proactive grant AQuS, Project No. 640800, and by Provincia Autonoma di Trento, partially through the project On silicon chip quantum optics for quantum computing and secure communications (SiQuro). H.M.P was also supported by the EC through the H2020 Marie Sklodowska-Curie Action, Individual Fellowship Grant No. 656093 SynOptic.
\end{acknowledgements}

\appendix
\section{Chern number in the parameter space of $\mathbf{r}_0$}
\label{sec:app}

In this appendix, we calculate the Chern number $\tilde{\mathcal{C}}$ in the parameter space of the trap center position $\mathbf{r}_0$. As in the main text, the flux in the underlying real-space Harper-Hofstadter model takes the form $\alpha = p/q$, where $p$ and $q$ are coprime. In the following, we assume $|\alpha| < 1$ and $q > 0$ without loss of generality. As discussed in~\cite{Ozawa:2015}, in the small trap limit, one effectively obtains a tight-binding model in momentum space, given by the following form:
\begin{align}
	\mathcal{H}^\mathrm{M}_\mathrm{TB}
	=
	-J^\prime
	&\sum_{\nu = 0}^{q-1}
	\left[
	\alpha^\dagger_{\nu+1}\alpha_\nu + \alpha^\dagger_\nu \alpha_{\nu+1}
	\right.
	\notag \\
	&\left.
	+ 2 \cos (2\pi x_0/q + 2\pi \mathcal{C}/q) \alpha^\dagger_\nu \alpha_\nu
	\right],
\end{align}
where $\mathcal{C}$ is the Chern number of the real-space Hofstadter lattice with a flux $\alpha = p/q$, and $\alpha_\nu$ is the annihilation operator of a particle in the Wannier state localized in momentum space at momentum $(0, 2\pi \nu/q)$. The dependence on $y_0$ enters through the boundary condition $\alpha_q = e^{i2\pi y_0} \alpha_0$. To make the $y_0$ dependence more explicit in the Hamiltonian, we now perform the transformation $\tilde{\alpha}_\nu \equiv e^{-i2\pi y_0 \nu/q} \alpha_\nu$. Then, the Hamiltonian is
\begin{align}
	\mathcal{H}^\mathrm{M}_\mathrm{TB}
	=
	-J^\prime
	&\sum_{\nu = 0}^{q-1}
	\left[
	e^{-i2\pi y_0/q} \tilde{\alpha}^\dagger_{\nu+1}\tilde{\alpha}_\nu + e^{i2\pi y_0/q}\tilde{\alpha}^\dagger_\nu \tilde{\alpha}_{\nu+1}
	\right.
	\notag \\
	&\left.
	+ 2 \cos (2\pi x_0/q + 2\pi \mathcal{C}\nu/q) \tilde{\alpha}^\dagger_\nu \tilde{\alpha}_\nu
	\right] \label{momHH}
\end{align}
with $\tilde{\alpha}_q = \tilde{\alpha}_0$.
Diagonalizing this Hamiltonian, one obtains the energy band structure in the parameter space of the trap center $\mathbf{r}_0$, and from the eigenstates, one can calculate the Berry curvature and the Chern number in $\mathbf{r}_0$-space.

The Chern number in $\mathbf{r}_0$ space can be, however, obtained without an explicit calculation by noticing the following similarity with the real-space Harper-Hofstadter model. For the real-space (untrapped) Harper-Hofstadter model with a flux through a plaquette $\alpha^\prime = p^\prime/q$ and the hopping amplitude $J^\prime$, the Hamiltonian is diagonal in $k_x$ and $k_y$, which are momenta along the $x$ and $y$ directions, and takes the form~\cite{Bernevig:Book}
\begin{align}
	\mathcal{H}_{k_x, k_y}
	=
	&-J^\prime
	\sum_{\nu=0}^{q-1}
	\left[
	e^{-ik_y}c_{k_x + 2\pi \alpha^\prime (\nu+1),k_y}^\dagger c_{k_x + 2\pi \alpha^\prime \nu,k_y}
	\right.
	\notag \\
	&\left.+
	e^{ik_y}c_{k_x + 2\pi \alpha^\prime \nu,k_y}^\dagger c_{k_x + 2\pi \alpha^\prime (\nu+1),k_y}
	\right.
	\notag \\
	&\left.+
	2\cos (k_x + 2\pi \alpha^\prime \nu) c_{k_x + 2\pi \alpha^\prime \nu,k_y}^\dagger c_{k_x + 2\pi \alpha^\prime \nu,k_y}
	\right] \label{realHH}
\end{align}
for a given $k_x$ and $k_y$, where $(k_x, k_y)$ lie within the magnetic Brillouin zone $0 \le k_x \le 2\pi/q$ and $0 \le k_y \le 2\pi$, and $c_{k_x,k_y}$ is the annihilator of a particle with a crystal momentum $(k_x, k_y)$.
Diagonalizing $\mathcal{H}_{k_x, k_y}$ one obtains the band structure and the Berry curvature in the Brillouin zone.
Comparing (\ref{momHH}) and (\ref{realHH}), one notices a clear similarity. In particular, the two Hamiltonians are the same once we make correspondence:
\begin{align}
	2\pi x_0/q &\longleftrightarrow k_x, \notag \\
	2\pi y_0/q &\longleftrightarrow k_y, \notag \\
	\mathcal{C}/q &\longleftrightarrow \alpha^\prime = p^\prime/q, \notag \\
	\tilde{\alpha}_{\nu} &\longleftrightarrow c_{k_x + 2\pi \alpha \nu,k_y}.
\end{align}
This implies that the Chern number in $\mathbf{r}_0$ space calculated from (\ref{momHH}) is equal to the Chern number of the ordinary Harper-Hofstadter model with the flux $\alpha^\prime = \mathcal{C}/q$.

For the Harper-Hofstadter model, there is a Diophantine equation which gives the Chern number of any band~\cite{Thouless:1982,Bernevig:Book}. The Diophantine equation applied to the lowest band of the Harper-Hofstadter model with flux $\alpha = p/q$ tells that the Chern number $\mathcal{C}$ is the unique integer in the range $-q/2 \le \mathcal{C} \le q/2$ which satisfies
\begin{align}
	1 = qs - p\mathcal{C}, \label{dio1}
\end{align}
where $s$ is an integer.
As derived above, the Chern number in $\mathbf{r}_0$ space is equal to the Chern number of an ordinary Harper-Hofstadter model with flux $\mathcal{C}/q$. This means that the Chern number of the lowest band in $\mathbf{r}_0$ space, $\tilde{\mathcal{C}}$, satisfies
\begin{align}
	1 = qs - \mathcal{C} \tilde{\mathcal{C}}. \label{dio2}
\end{align}
Comparing (\ref{dio1}) and (\ref{dio2}), one obtains that when $|p| \le q/2$, $\tilde{\mathcal{C}} = p$, when $p > q/2$, $\tilde{\mathcal{C}} = p-q$, and when $p < -q/2$, $\tilde{\mathcal{C}} = p+q$.
In particular, when $\alpha = 1/q$, $\tilde{\mathcal{C}} = 1$ as used in the main text.

Even though the derivation given here is restricted to the small trap limit, the result for the Chern number $\tilde{\mathcal{C}}$ remains correct also for larger trap strengths, as long as the energy gap in $\mathbf{r}_0$ remains open.

\end{document}